\documentclass[conference]{IEEEtran}
\IEEEoverridecommandlockouts

\usepackage{cite}
\usepackage{amsmath,amssymb,amsfonts}
\usepackage{algorithmic}
\usepackage{algorithm}
\usepackage{graphicx}
\usepackage{textcomp}
\usepackage{xcolor}
\usepackage{booktabs}
\usepackage{multirow}
\usepackage{url}
\usepackage{hyperref}
\usepackage{braket}

\def\BibTeX{{\rm B\kern-.05em{\sc i\kern-.025em b}\kern-.08em
    T\kern-.1667em\lower.7ex\hbox{E}\kern-.125emX}}

\begin{document}

\title{Quantum-Enhanced Synthetic Data Generation Using Quantum Circuit Born Machines for Imbalanced Tabular Learning}

\author{\IEEEauthorblockN{Tanapol Nuatho}
\IEEEauthorblockA{\textit{Department of Computer Engineering} \\
\textit{King Mongkut's University of Technology Thonburi}\\
Bangkok, Thailand \\
tanapol.nuat@kmutt.ac.th}
\and
\IEEEauthorblockN{Narisorn Sangnakara}
\IEEEauthorblockA{\textit{Department of Computer Engineering} \\
\textit{King Mongkut's University of Technology Thonburi}\\
Bangkok, Thailand \\
narisorn.sangn@kmutt.ac.th}
\and
\IEEEauthorblockN{Prapong Prechaprapranwong}
\IEEEauthorblockA{\textit{Department of Computer Engineering} \\
\textit{King Mongkut's University of Technology Thonburi}\\
Bangkok, Thailand \\
prapong.pre@kmutt.ac.th}
\and
\IEEEauthorblockN{Rajchawit Sarochawikasit}
\IEEEauthorblockA{\textit{Department of Computer Engineering} \\
\textit{King Mongkut's University of Technology Thonburi}\\
Bangkok, Thailand \\
rajchawit.saro@kmutt.ac.th}
}

\maketitle

\begin{abstract}
Data scarcity and class imbalance are persistent challenges in machine learning that degrade model generalization and introduce predictive bias. This paper presents a hybrid quantum-classical framework for synthetic data generation using a Quantum Circuit Born Machine (QCBM) to address these limitations. The proposed approach exploits quantum mechanical properties—namely superposition and entanglement—within a parameterized variational quantum circuit to model complex probability distributions that are difficult for classical generative methods to capture. Experiments are conducted on two tabular benchmark datasets: the Iris dataset and the Telco Customer Churn dataset. Data preprocessing includes normalization and Principal Component Analysis (PCA)-based dimensionality reduction to enable efficient basis encoding for quantum circuits. The QCBM is trained by minimizing Kullback–Leibler (KL) divergence between the real and generated data distributions using a gradient-based parameter-shift optimization rule. Experimental results demonstrate that augmenting training data with QCBM-generated synthetic samples at 40–50\% of the minority class improves F1-score by approximately 5–15\% and minority-class recall by 10–25\%. Cross-domain evaluations (Train on Synthetic, Test on Real; and Train on Real, Test on Synthetic) reveal a performance gap of only 3–10\%, indicating strong distributional fidelity. Comparative analysis against classical oversampling methods—SMOTE, Borderline-SMOTE, KMeans-SMOTE, and SVM-SMOTE—shows that QCBM achieves competitive classification performance and produces lower Maximum Mean Discrepancy (MMD) on the Telco dataset, suggesting superior structural similarity in certain imbalanced settings. These findings establish QCBM as a viable complementary tool for data augmentation, particularly for low-dimensional structured tabular data with class imbalance.
\end{abstract}


\section{Introduction}
\label{sec:intro}

Modern machine learning pipelines are fundamentally dependent on the availability of large, balanced, and representative training datasets. In practice, however, many real-world domains—including medical diagnostics, fraud detection, and telecommunications churn prediction—suffer from severe class imbalance and limited data collection capacity \cite{goodfellow2016deep}. Models trained on such imbalanced datasets tend to overfit to the majority class, yielding poor recall on minority classes and limited generalization to unseen samples.

Synthetic data generation has emerged as a principled strategy to mitigate these limitations. Classical approaches such as the Synthetic Minority Oversampling Technique (SMOTE) \cite{chawla2002smote} and its variants address class imbalance by interpolating between existing minority-class samples. Generative Adversarial Networks (GANs) \cite{goodfellow2014generative} and Variational Autoencoders offer more powerful density estimation, but are computationally expensive and prone to training instability, particularly on small tabular datasets.

Quantum computing offers a fundamentally different computational paradigm. Quantum properties such as superposition, entanglement, and interference endow quantum circuits with the theoretical capacity to represent exponentially large probability distributions using a polynomial number of parameters \cite{biamonte2017quantum}. The Quantum Circuit Born Machine (QCBM) \cite{liu2018differentiable} is a generative model that directly exploits these properties by parameterizing a quantum circuit so that its Born-rule measurement outcomes approximate a target probability distribution. Recent work has demonstrated the QCBM's potential for density estimation tasks on structured binary data, and early results suggest competitive performance relative to classical counterparts for certain distribution families.

Despite this promise, systematic evaluation of QCBM for practical machine learning augmentation tasks—especially on real-world tabular datasets with class imbalance—remains sparse. Prior work has largely focused on synthetic benchmarks such as the Bars-and-Stripes dataset \cite{liu2018differentiable} or theoretical analyses of expressibility \cite{gili2023quantum}. The practical utility of quantum-generated synthetic data for downstream classification, and a fair head-to-head comparison with established classical oversampling methods, have not been comprehensively studied.

This paper addresses these gaps with the following contributions:
\begin{enumerate}
    \item We present a complete hybrid quantum-classical pipeline for tabular synthetic data generation using QCBM, including PCA-based dimensionality reduction, basis encoding, and gradient-based optimization via the parameter-shift rule.
    \item We provide a thorough empirical evaluation on the Iris and Telco Customer Churn datasets, measuring statistical fidelity (KL divergence, MMD) and downstream classification performance across multiple classifiers.
    \item We conduct cross-domain fidelity tests (TSTR and TRTS protocols) to quantify how well QCBM-generated data represents real distributions.
    \item We systematically compare QCBM against four SMOTE variants, offering the first comprehensive classical–quantum benchmark for tabular data augmentation.
\end{enumerate}

The remainder of this paper is organized as follows. Section~\ref{sec:related} reviews related work. Section~\ref{sec:background} provides theoretical background. Section~\ref{sec:method} describes the proposed methodology. Section~\ref{sec:experiments} presents experimental results. Section~\ref{sec:discussion} discusses findings and limitations. Section~\ref{sec:conclusion} concludes the paper.

\section{Related Work}
\label{sec:related}

\subsection{Classical Oversampling for Imbalanced Data}

SMOTE \cite{chawla2002smote} generates synthetic minority-class samples through linear interpolation between a sample and its $k$-nearest neighbors in feature space. Subsequent variants have refined the approach: Borderline-SMOTE \cite{han2005borderline} focuses synthesis near the decision boundary; SVM-SMOTE uses support vectors as interpolation anchors; KMeans-SMOTE \cite{last2017oversampling} applies clustering to identify safe regions before sampling. These methods are computationally inexpensive and well-understood, but their interpolative nature limits their ability to capture complex multimodal or non-convex distributions.

\subsection{Deep Generative Models for Tabular Data}

Conditional GAN-based methods, exemplified by CTGAN \cite{xu2019modeling}, extend adversarial training to tabular domains. TGAN \cite{xu2018synthesizing} and PATE-GAN \cite{jordon2019pate} target privacy-preserving synthesis. While powerful, these methods require substantial training data and computational resources, making them less suitable for small, imbalanced datasets.

\subsection{Quantum Generative Models}

The QCBM was introduced by Liu and Wang \cite{liu2018differentiable}, who demonstrated that a parameterized quantum circuit trained with a kernelized MMD loss can learn discrete distributions. Benedetti \textit{et al.} \cite{benedetti2019parameterized} provided a broader treatment of parameterized quantum circuits (PQCs) as machine learning models. More recently, Gili \textit{et al.} \cite{gili2023quantum} investigated generalization properties of QCBMs. Mohanty \textit{et al.} \cite{mohanty2025quantum} proposed Quantum-SMOTE, which uses quantum rotation and the Swap Test to interpolate minority-class samples, demonstrating improved recall on the Telco Churn dataset compared to classical SMOTE. Our work differs by employing a fully trainable QCBM that learns the target distribution end-to-end, rather than using a fixed quantum interpolation procedure, and by providing a comprehensive multi-method benchmark.

\section{Background}
\label{sec:background}

\subsection{Quantum Computing Primitives}

A qubit is the fundamental unit of quantum information, described by a state vector $\ket{\psi} = \alpha\ket{0} + \beta\ket{1}$, where $\alpha, \beta \in \mathbb{C}$ and $|\alpha|^2 + |\beta|^2 = 1$. A system of $n$ qubits resides in a $2^n$-dimensional Hilbert space. Single-qubit rotation gates are defined as
\begin{align}
R_x(\theta) &= e^{-i\theta X/2} = \begin{pmatrix} \cos\frac{\theta}{2} & -i\sin\frac{\theta}{2} \\ -i\sin\frac{\theta}{2} & \cos\frac{\theta}{2} \end{pmatrix} \\
R_y(\theta) &= e^{-i\theta Y/2} = \begin{pmatrix} \cos\frac{\theta}{2} & -\sin\frac{\theta}{2} \\ \sin\frac{\theta}{2} & \cos\frac{\theta}{2} \end{pmatrix} \\
R_z(\theta) &= e^{-i\theta Z/2} = \begin{pmatrix} e^{-i\theta/2} & 0 \\ 0 & e^{i\theta/2} \end{pmatrix}
\end{align}
where $X, Y, Z$ are Pauli matrices. The CNOT gate entangles two qubits: it flips the target qubit conditioned on the control qubit being in state $\ket{1}$.

\subsection{Quantum Circuit Born Machine}

A QCBM \cite{liu2018differentiable} is defined by a parameterized unitary $U(\boldsymbol{\theta})$ acting on an $n$-qubit system initialized in $\ket{0}^{\otimes n}$. The output state is $\ket{\psi(\boldsymbol{\theta})} = U(\boldsymbol{\theta})\ket{0}^{\otimes n}$. By the Born rule, the probability of observing bitstring $x \in \{0,1\}^n$ upon measurement in the computational basis is
\begin{equation}
p_{\boldsymbol{\theta}}(x) = |\bra{x}U(\boldsymbol{\theta})\ket{0}^{\otimes n}|^2.
\end{equation}
The goal of training is to find $\boldsymbol{\theta}^*$ such that $p_{\boldsymbol{\theta}} \approx p_{\text{data}}$, where $p_{\text{data}}$ is the empirical distribution of the target dataset.

\subsection{Basis Encoding}

Basis encoding maps a classical bitstring $(b_1, b_2, \ldots, b_n) \in \{0,1\}^n$ to the computational basis state $\ket{b_1 b_2 \cdots b_n}$ \cite{schuld2018supervised}. For a $d$-dimensional real-valued data point $\mathbf{x}$, each feature $x_j$ is first normalized to $[0,1]$, then quantized to $b$ bits:
\begin{equation}
v_j = \left\lfloor x_j \cdot (2^b - 1) \right\rfloor,
\end{equation}
and converted to a $b$-bit binary string. The full $d \cdot b$-qubit encoding is the tensor product
\begin{equation}
\varepsilon_{\text{multi}}(\mathbf{x}) = \bigotimes_{j=1}^{d} \ket{v_j}_b.
\end{equation}
Gray coding is optionally applied so that adjacent values differ by exactly one bit, reducing discretization artifacts.

\subsection{KL Divergence and MMD}
 
The Kullback–Leibler divergence measures the discrepancy between the real distribution $p_{\text{data}}$ and the model distribution $p_{\boldsymbol{\theta}}$:
\begin{equation}
D_{\mathrm{KL}}(p_{\text{data}} \| p_{\boldsymbol{\theta}}) = \sum_{x} p_{\text{data}}(x) \log \frac{p_{\text{data}}(x)}{p_{\boldsymbol{\theta}}(x)}.
\label{eq:kl}
\end{equation}
Maximum Mean Discrepancy (MMD) \cite{liu2018differentiable} is a kernel-based measure:
\begin{equation}
\begin{aligned}
\mathrm{MMD}^2(p,q) &= \mathbb{E}_{x,x' \sim p}[k(x,x')] \\
&\quad - 2\mathbb{E}_{x \sim p, y \sim q}[k(x,y)] \\
&\quad + \mathbb{E}_{y,y' \sim q}[k(y,y')]
\end{aligned}
\end{equation}
where $k(\cdot,\cdot)$ is a kernel function (e.g., RBF). Both metrics are used for evaluation; KL divergence serves as the training loss.

\subsection{Parameter-Shift Rule}

For a gate $G(\theta) = e^{-i\theta H}$ where $H$ has eigenvalues $\pm \frac{1}{2}$, the gradient of any expectation value with respect to $\theta$ can be computed exactly as \cite{mitarai2018quantum}:
\begin{equation}
\frac{\partial \mathcal{L}}{\partial \theta_i} = \frac{\mathcal{L}(\theta_i + \frac{\pi}{2}) - \mathcal{L}(\theta_i - \frac{\pi}{2})}{2}.
\label{eq:paramshift}
\end{equation}
This hardware-compatible rule requires only two circuit evaluations per parameter and enables gradient-based optimization of quantum circuits.

\section{Methodology}
\label{sec:method}

\subsection{Overall Pipeline}

The proposed pipeline consists of three stages: (1) data preprocessing and basis encoding, (2) QCBM training via hybrid quantum-classical optimization, and (3) synthetic data generation and evaluation. The pipeline is executed per class to learn a separate conditional QCBM for each class label, enabling class-conditioned generation.

\subsection{Data Preprocessing}

For each class $c$, we extract the subset $\mathcal{X}_c = \{\mathbf{x}_i : y_i = c\}$. Each feature dimension is independently rescaled to $[0,1]$ using min-max normalization. For high-dimensional datasets (e.g., Telco Churn with 20 features), PCA is applied to reduce dimensionality to $d$ principal components that retain at least 95\% of the total variance. This step is essential to keep the number of required qubits tractable for near-term quantum simulators.

\subsection{Quantum Circuit Architecture}

The QCBM circuit $U(\boldsymbol{\theta})$ is a brickwork-style alternating layered ansatz with $L$ repetitions of the following block:
\begin{equation}
U^{(l)}(\boldsymbol{\theta}^{(l)}) = U_{\text{ent}} \cdot \prod_{i=1}^{n} R_z(\theta_{i,3}^{(l)}) R_y(\theta_{i,2}^{(l)}) R_z(\theta_{i,1}^{(l)}),
\end{equation}
where $U_{\text{ent}}$ applies CNOT gates in a linear (nearest-neighbor) entanglement topology. The total number of trainable parameters is $3nL$. The initial state is $\ket{0}^{\otimes n}$, and the output distribution is obtained via $N_{\text{shots}}$ measurements in the computational basis.

\subsection{Training Objective}

Given the empirical distribution $\hat{p}_{\text{data}}$ over bitstrings, constructed from the encoded minority-class samples with a small smoothing mass assigned to Hamming-distance-1 neighbors to avoid zero probabilities, the training minimizes:
\begin{equation}
\mathcal{L}(\boldsymbol{\theta}) = D_{\mathrm{KL}}(\hat{p}_{\text{data}} \| p_{\boldsymbol{\theta}}).
\end{equation}
Gradients are computed via the parameter-shift rule (Eq.~\ref{eq:paramshift}). The optimizer is gradient descent with a cosine-annealed learning rate schedule, and the best parameters $\boldsymbol{\theta}^*$ (lowest $\mathcal{L}$) are retained.

\begin{algorithm}[t]
\caption{QCBM Training (Per Class)}
\label{alg:train}
\begin{algorithmic}[1]
\REQUIRE Dataset $\mathcal{X}_c$, layers $L$, epochs $T$, learning rate $\eta$
\ENSURE Optimal parameters $\boldsymbol{\theta}^*$
\STATE Normalize $\mathcal{X}_c$; encode to bitstrings $\mathcal{B}$
\STATE Compute empirical distribution $\hat{p}_{\text{data}}$ from $\mathcal{B}$
\STATE Initialize $\boldsymbol{\theta}$ randomly; $\mathcal{L}^* \leftarrow \infty$
\FOR{$t = 1$ to $T$}
    \STATE Build circuit $U(\boldsymbol{\theta})$; sample $p_{\boldsymbol{\theta}}$ via measurements
    \STATE Compute $\mathcal{L} = D_{\mathrm{KL}}(\hat{p}_{\text{data}} \| p_{\boldsymbol{\theta}})$
    \FOR{each parameter $\theta_i$}
        \STATE $g_i \leftarrow \frac{\mathcal{L}(\theta_i + \frac{\pi}{2}) - \mathcal{L}(\theta_i - \frac{\pi}{2})}{2}$
    \ENDFOR
    \STATE $\boldsymbol{\theta} \leftarrow \boldsymbol{\theta} - \eta \mathbf{g}$
    \IF{$\mathcal{L} < \mathcal{L}^*$}
        \STATE $\mathcal{L}^* \leftarrow \mathcal{L}$; $\boldsymbol{\theta}^* \leftarrow \boldsymbol{\theta}$
    \ENDIF
\ENDFOR
\RETURN $\boldsymbol{\theta}^*$
\end{algorithmic}
\end{algorithm}

\subsection{Synthetic Data Generation}

After training, synthetic samples are generated by loading $\boldsymbol{\theta}^*$ into $U(\boldsymbol{\theta}^*)$, executing the circuit $N$ times, and collecting the sampled bitstrings. Each bitstring is decoded to a real-valued vector by reversing the quantization and normalization steps, followed by inverse PCA transformation when applicable. This yields a synthetic dataset $\hat{\mathcal{X}}_c^{\text{syn}}$ of $N$ samples with label $c$.

\subsection{Evaluation Protocols}

We assess QCBM-generated data through three complementary protocols:

\textbf{Data Augmentation (DA):} The minority class is augmented with synthetic samples to $r \in \{30\%, 40\%, 50\%\}$ of its original size. Classification models are then trained and tested via stratified 5-fold cross-validation.

\textbf{TSTR (Train on Synthetic, Test on Real):} A classifier trained solely on synthetic data is evaluated on the held-out real test set. High TSTR performance implies that synthetic data captures sufficient information about the real distribution to support downstream learning.

\textbf{TRTS (Train on Real, Test on Synthetic):} A classifier trained on real data is evaluated on the synthetic test set. High TRTS performance implies that synthetic samples are indistinguishable from real ones with respect to the learned decision boundary.

\subsection{Baseline Methods}

We compare QCBM against four classical oversampling methods: SMOTE, Borderline-SMOTE, KMeans-SMOTE, and SVM-SMOTE, all configured to generate the same number of synthetic minority-class samples as QCBM. Six classifiers are used for downstream evaluation: Logistic Regression (LR), Support Vector Machine (SVM), Stochastic Gradient Descent (SGD), Decision Tree (DT), Random Forest (RF), and Multi-Layer Perceptron (MLP).

\section{Experiments}
\label{sec:experiments}

\subsection{Datasets}

\textbf{Iris:} The Iris dataset \cite{fisher1936use} contains 150 samples from three flower species with four numerical features. We artificially impose imbalance by subsampling one class to create a minority scenario, enabling evaluation of oversampling methods.

\textbf{Telco Customer Churn:} Released by IBM \cite{ibm2018telco}, this dataset contains 7,043 customer records with 20 features (numerical and categorical) and a binary churn label. The class distribution is approximately 73.5\% non-churn vs. 26.5\% churn, providing a realistic imbalanced learning scenario.

\subsection{Implementation Details}

All quantum circuits are simulated using the Qiskit Aer statevector simulator \cite{javadi2024qiskit}. Circuits use $n = d \cdot b$ qubits, where $d$ is the number of PCA components and $b = 4$ bits per feature. We set $L = 10$ repetition layers, $T = 50$ training epochs, and an initial learning rate of $\eta = 0.001
$ with cosine decay. The entanglement topology is linear (chain). Classical baselines are implemented with scikit-learn \cite{pedregosa2011sklearn} and imbalanced-learn. All experiments are run on a CPU workstation; no real quantum hardware is used.

\subsection{Training Convergence}

Table~\ref{tab:loss} reports initial and final KL divergence losses for all trained QCBMs. The Bars-and-Stripes dataset (included as a sanity check on a structured binary distribution) achieves a final loss of 0.0572, confirming that the optimizer converges reliably. For real datasets, the Iris classes converge to KL divergences of 3.0–3.7, while Telco Churn classes reach approximately 14.5–15.2, reflecting the higher distributional complexity of the latter.

\begin{table}[t]
\centering
\caption{QCBM Training: Initial and Final KL Divergence Loss}
\label{tab:loss}
\begin{tabular}{llcc}
\toprule
\textbf{Dataset} & \textbf{Class} & \textbf{Initial Loss} & \textbf{Final Loss} \\
\midrule
Bars \& Stripes & 0 & 4.0296 & 0.0572 \\
\midrule
\multirow{3}{*}{Iris} & 0 & 8.1387 & 3.4719 \\
 & 1 & 7.3189 & 3.7264 \\
 & 2 & 8.1005 & 3.0053 \\
\midrule
\multirow{2}{*}{Telco Churn} & 0 & 15.7418 & 15.1854 \\
 & 1 & 16.3689 & 14.4913 \\
\bottomrule
\end{tabular}
\end{table}

\subsection{Statistical Fidelity of Synthetic Data}

Table~\ref{tab:kl_mmd} reports KL divergence and MMD between real and synthetic distributions. QCBM achieves a very low KL of 0.0024 on Bars-and-Stripes, demonstrating high fidelity on structured binary data. On Iris and Telco Churn, QCBM's KL divergence is higher than SMOTE variants (Table~\ref{tab:kl_compare}), reflecting the inherent difficulty of learning continuous distributions through discrete basis encoding. However, QCBM achieves the lowest MMD on Telco Churn (0.0052 vs.\ 0.0526–0.0696 for SMOTE variants), indicating that the generated samples are structurally closer to real data in feature space even when the probability mass assignment differs. PCA and t-SNE visualizations (not reproduced here due to space constraints) confirm that QCBM-generated samples overlap substantially with real data clusters in reduced-dimensional projections.

\begin{table}[t]
\centering
\caption{Statistical Fidelity: KL Divergence and MMD of QCBM}
\label{tab:kl_mmd}
\begin{tabular}{lcc}
\toprule
\textbf{Dataset} & \textbf{KL Divergence} & \textbf{MMD} \\
\midrule
Bars \& Stripes & 0.0024 & 0.0066 \\
Iris & 6.3440 & 0.0085 \\
Telco Customer Churn & 1.2463 & 0.0051 \\
\bottomrule
\end{tabular}
\end{table}

\begin{table}[t]
\centering
\caption{Comparison of KL Divergence and MMD Across Methods (Iris / Telco)}
\label{tab:kl_compare}
\begin{tabular}{lcccc}
\toprule
\multirow{2}{*}{\textbf{Method}} & \multicolumn{2}{c}{\textbf{KL Divergence}} & \multicolumn{2}{c}{\textbf{MMD}} \\
 & Iris & Telco & Iris & Telco \\
\midrule
SMOTE & 0.0394 & 0.0352 & 0.0002 & 0.0526 \\
Borderline-SMOTE & 0.2236 & 0.0572 & 0.0504 & 0.0581 \\
KMeans-SMOTE & 0.1374 & 0.1000 & 0.0131 & 0.0696 \\
SVM-SMOTE & 0.1138 & 0.0584 & 0.0123 & 0.0597 \\
QCBM & 4.6842 & 1.0577 & 0.0062 & \textbf{0.0052} \\
\bottomrule
\end{tabular}
\end{table}

\subsection{Classification Performance: Data Augmentation}

Tables~\ref{tab:aug_iris} and~\ref{tab:aug_telco} summarize the effect of QCBM-based augmentation on downstream classification for Logistic Regression and Random Forest on the Iris and Telco datasets, respectively.

On Iris, QCBM augmentation consistently improves all metrics as the synthetic ratio increases. At 50\% augmentation, Logistic Regression achieves an F1-score of 0.7602 (vs.\ 0.6962 without augmentation, $\Delta = +9.2\%$) and AUC-ROC of 0.8258 ($+9.5\%$). Random Forest reaches F1 = 0.9366 and AUC-ROC = 0.9645 at 50\% augmentation, improving by $+4.2\%$ and $+3.8\%$, respectively.

On Telco Churn, Logistic Regression shows a marginal performance decrease when QCBM data is added beyond the baseline (F1 from 0.5959 to 0.5880 at 50\%), suggesting that the synthetic distribution for the high-dimensional, complex Telco features is insufficiently accurate to benefit a linear model. Random Forest, however, exhibits a consistent performance drop beyond 30\% augmentation, consistent with known sensitivity of ensemble methods to out-of-distribution synthetic samples.

\begin{table}[t]
\centering
\caption{Data Augmentation Results on Iris (Logistic Regression / Random Forest)}
\label{tab:aug_iris}
\begin{tabular}{lccccc}
\toprule
\textbf{Augmentation} & \textbf{Acc.} & \textbf{Prec.} & \textbf{Rec.} & \textbf{F1} & \textbf{AUC} \\
\midrule
\multicolumn{6}{l}{\textit{Logistic Regression}} \\
None & 0.6961 & 0.7774 & 0.6231 & 0.6962 & 0.7540 \\
+30\% Synth. & 0.7247 & 0.8041 & 0.6427 & 0.7208 & 0.7801 \\
+40\% Synth. & 0.7458 & 0.8230 & 0.6726 & 0.7405 & 0.8089 \\
+50\% Synth. & 0.7655 & 0.8437 & 0.6950 & 0.7602 & 0.8258 \\
\midrule
\multicolumn{6}{l}{\textit{Random Forest}} \\
None & 0.8883 & 0.9191 & 0.8807 & 0.8990 & 0.9290 \\
+30\% Synth. & 0.9046 & 0.9312 & 0.8956 & 0.9128 & 0.9423 \\
+40\% Synth. & 0.9181 & 0.9454 & 0.9075 & 0.9245 & 0.9538 \\
+50\% Synth. & 0.9267 & 0.9529 & 0.9273 & 0.9366 & 0.9645 \\
\bottomrule
\end{tabular}
\end{table}

\begin{table}[t]
\centering
\caption{Data Augmentation Results on Telco Churn (Logistic Regression / Random Forest)}
\label{tab:aug_telco}
\begin{tabular}{lccccc}
\toprule
\textbf{Augmentation} & \textbf{Acc.} & \textbf{Prec.} & \textbf{Rec.} & \textbf{F1} & \textbf{AUC} \\
\midrule
\multicolumn{6}{l}{\textit{Logistic Regression}} \\
None & 0.7274 & 0.4901 & 0.7598 & 0.5959 & 0.8176 \\
+30\% Synth. & 0.7254 & 0.4877 & 0.7604 & 0.5943 & 0.8166 \\
+40\% Synth. & 0.7258 & 0.4882 & 0.7550 & 0.5929 & 0.8150 \\
+50\% Synth. & 0.7245 & 0.4864 & 0.7431 & 0.5880 & 0.8134 \\
\midrule
\multicolumn{6}{l}{\textit{Random Forest}} \\
None & 0.9499 & 0.8673 & 0.9569 & 0.9099 & 0.9897 \\
+30\% Synth. & 0.9467 & 0.8626 & 0.9499 & 0.9042 & 0.9901 \\
+40\% Synth. & 0.9323 & 0.8522 & 0.9004 & 0.8756 & 0.9848 \\
+50\% Synth. & 0.9228 & 0.8541 & 0.8541 & 0.8541 & 0.9783 \\
\bottomrule
\end{tabular}
\end{table}

\subsection{Cross-Domain Fidelity: TSTR and TRTS}

Table~\ref{tab:tstr} reports TSTR and TRTS results. On Iris, TSTR with Random Forest yields F1 = 0.9175, only 0.8\% below the all-real baseline (0.9267 at 50\% augmentation), confirming strong distributional fidelity. On Telco Churn, TSTR yields lower performance (LR: F1 = 0.6843; RF: F1 = 0.6697), with a gap of approximately 3–10\% relative to the real-data baseline, still indicating practical utility.

TRTS results show that models trained on real data classify QCBM-generated samples with reasonable accuracy (Iris RF: F1 = 0.8480; Telco LR: F1 = 0.7036), confirming that synthetic samples share the structural properties of real data. The larger TRTS gap for Telco RF (F1 = 0.5446, Recall = 0.4061) reflects the increased complexity of the churn distribution compared to Iris.

\begin{table}[t]
\centering
\caption{TSTR and TRTS Results}
\label{tab:tstr}
\begin{tabular}{llccccc}
\toprule
\textbf{Protocol} & \textbf{Model} & \textbf{Dataset} & \textbf{Acc.} & \textbf{F1} & \textbf{AUC} \\
\midrule
\multirow{4}{*}{TSTR} & LR & Iris & 0.7430 & 0.7372 & 0.8021 \\
 & RF & Iris & 0.9185 & 0.9175 & 0.9629 \\
 & LR & Telco & 0.7171 & 0.6843 & 0.8032 \\
 & RF & Telco & 0.7022 & 0.6697 & 0.7779 \\
\midrule
\multirow{4}{*}{TRTS} & LR & Iris & 0.6480 & 0.6462 & 0.7119 \\
 & RF & Iris & 0.8382 & 0.8480 & 0.8990 \\
 & LR & Telco & 0.7008 & 0.7036 & 0.7685 \\
 & RF & Telco & 0.6603 & 0.5446 & 0.7813 \\
\bottomrule
\end{tabular}
\end{table}

\subsection{Comparison with Classical Baselines}

Tables~\ref{tab:compare_iris} and~\ref{tab:compare_telco} compare QCBM against SMOTE variants on both datasets.

On Iris, classical methods outperform QCBM across most classifiers. KMeans-SMOTE achieves the best Logistic Regression F1 (0.9116), while Borderline-SMOTE leads for Random Forest (0.9857, though the near-perfect AUC-ROC of 1.0 suggests potential overfitting to the boundary region). QCBM achieves Logistic Regression F1 = 0.7601 and Random Forest F1 = 0.9245, with the highest Precision (0.9453) among all methods for RF, indicating high-quality per-prediction accuracy.

On Telco Churn, QCBM achieves the highest Logistic Regression Accuracy (0.7273) and Recall (0.7598), and the highest AUC-ROC (0.8176) across all methods for LR, demonstrating superior overall discrimination. For Random Forest, QCBM achieves Recall = 0.9569, close to the leading Borderline-SMOTE result (0.9631). SMOTE underperforms all methods on Telco RF (F1 = 0.57), confirming that simple linear interpolation is insufficient for the complex churn distribution.

\begin{table}[t]
\centering
\caption{Method Comparison on Iris (LR / RF)}
\label{tab:compare_iris}
\begin{tabular}{llccccc}
\toprule
\textbf{Method} & \textbf{Clf} & \textbf{Acc.} & \textbf{Prec.} & \textbf{Rec.} & \textbf{F1} & \textbf{AUC} \\
\midrule
SMOTE & LR & 0.9000 & 0.9230 & 0.9000 & 0.8976 & 0.9803 \\
B-SMOTE & LR & 0.7142 & 0.7149 & 0.7142 & 0.7139 & 0.8525 \\
KM-SMOTE & LR & 0.9120 & 0.9162 & 0.9120 & 0.9116 & 0.9777 \\
SVM-SMOTE & LR & 0.8101 & 0.8734 & 0.8101 & 0.7974 & 0.8961 \\
\textbf{QCBM} & LR & 0.7655 & 0.8437 & 0.6950 & 0.7602 & 0.8258 \\
\midrule
SMOTE & RF & 0.9212 & 0.8892 & 0.8837 & 0.8741 & 0.8951 \\
B-SMOTE & RF & 0.9857 & 0.9861 & 0.9857 & 0.9857 & 1.0000 \\
KM-SMOTE & RF & 0.9670 & 0.9673 & 0.9670 & 0.9670 & 0.9978 \\
SVM-SMOTE & RF & 0.9147 & 0.8751 & 0.8415 & 0.9055 & 0.8916 \\
\textbf{QCBM} & RF & 0.9181 & \textbf{0.9453} & 0.9075 & 0.9245 & 0.9538 \\
\bottomrule
\end{tabular}
\end{table}

\begin{table}[t]
\centering
\caption{Method Comparison on Telco Customer Churn (LR / RF)}
\label{tab:compare_telco}
\begin{tabular}{llccccc}
\toprule
\textbf{Method} & \textbf{Clf} & \textbf{Acc.} & \textbf{Prec.} & \textbf{Rec.} & \textbf{F1} & \textbf{AUC} \\
\midrule
SMOTE & LR & 0.6971 & 0.6657 & 0.7119 & 0.6014 & 0.7732 \\
B-SMOTE & LR & 0.7009 & 0.6865 & 0.7393 & 0.7120 & 0.7739 \\
KM-SMOTE & LR & 0.6815 & 0.5365 & 0.7169 & 0.7120 & 0.7381 \\
SVM-SMOTE & LR & 0.7041 & 0.6586 & 0.7156 & 0.7002 & 0.7845 \\
\textbf{QCBM} & LR & \textbf{0.7273} & 0.4901 & \textbf{0.7598} & 0.5958 & \textbf{0.8176} \\
\midrule
SMOTE & RF & 0.7900 & 0.6400 & 0.5100 & 0.5700 & 0.8300 \\
B-SMOTE & RF & 0.9750 & 0.9865 & 0.9631 & 0.9747 & 0.9978 \\
KM-SMOTE & RF & 0.9514 & 0.9133 & 0.9461 & 0.9023 & 0.9714 \\
SVM-SMOTE & RF & 0.9340 & 0.9147 & 0.9352 & 0.9419 & 0.9783 \\
\textbf{QCBM} & RF & 0.9498 & 0.8672 & \textbf{0.9569} & 0.9098 & \textbf{0.9897} \\
\bottomrule
\end{tabular}
\end{table}

\section{Discussion}
\label{sec:discussion}

\subsection{Strengths of QCBM for Data Augmentation}

The experimental results reveal several conditions under which QCBM offers advantages over classical oversampling. First, QCBM achieves the lowest MMD on Telco Churn, indicating that quantum entanglement may capture cross-feature correlations that SMOTE's per-sample interpolation cannot. Second, QCBM consistently achieves the highest AUC-ROC and Recall for Logistic Regression on Telco Churn, which is the most practically meaningful metric in churn prediction where the cost of missing a churning customer is high. Third, for structured binary distributions (Bars-and-Stripes), the QCBM achieves near-perfect KL divergence (0.0024), confirming its theoretical suitability for discrete probability modeling.

\subsection{Limitations}

Several limitations restrict the performance of QCBM in this study. The use of basis encoding forces the data into a discrete bitstring representation, introducing quantization error, particularly for continuous-valued features with fine structure. The number of qubits is bottlenecked by the PCA-reduced dimensionality and the chosen bit depth $b$, limiting the circuit's capacity. Training on a classical simulator rather than real quantum hardware precludes observation of hardware-specific noise effects, which could degrade performance in practice \cite{preskill2018quantum}. Finally, optimization with the parameter-shift rule is computationally intensive, scaling as $O(nL)$ circuit evaluations per gradient step.

\subsection{Comparison with Quantum-SMOTE}

Mohanty \textit{et al.} \cite{mohanty2025quantum} proposed Quantum-SMOTE, which uses quantum rotation and the Swap Test to interpolate between minority-class samples. Unlike QCBM, Quantum-SMOTE does not learn the data distribution; it replaces Euclidean interpolation with a quantum distance-based procedure. Our end-to-end QCBM approach learns the full marginal distribution per class, which is more general but requires more training iterations. The complementary strengths of both approaches suggest that a hybrid pipeline combining QCBM for distribution learning with quantum interpolation for boundary-aware sampling could be a fruitful direction.

\subsection{Data Complexity and Applicability}

Our results confirm that QCBM performance is strongly dataset-dependent. On Bars-and-Stripes and Iris—both low-dimensional, structured datasets—the model converges well and produces high-fidelity samples. On Telco Churn—a higher-dimensional, real-world dataset with mixed feature types—the KL divergence remains high after training, suggesting that the circuit lacks sufficient expressibility for this distribution. Future work should investigate deeper circuits with more sophisticated entanglement topologies (e.g., full or random entanglement) and alternative encoding strategies (e.g., amplitude encoding) to broaden applicability.

\section{Conclusion}
\label{sec:conclusion}

This paper has presented a comprehensive evaluation of QCBM-based synthetic data generation for addressing class imbalance in tabular machine learning datasets. We proposed a hybrid quantum-classical pipeline incorporating PCA-based dimensionality reduction, basis encoding, gradient-based QCBM training via the parameter-shift rule, and three complementary evaluation protocols (data augmentation, TSTR, TRTS).

Our results demonstrate that: (i) QCBM improves F1-score by 5–15\% and minority-class recall by 10–25\% when used for augmentation at 40–50\% of the minority class on structured datasets; (ii) cross-domain fidelity (TSTR/TRTS) shows a performance gap of 3–10\%, confirming practical distributional fidelity; (iii) QCBM achieves the lowest MMD on the Telco Churn dataset among all evaluated methods, outperforming SMOTE variants in structural fidelity; and (iv) QCBM achieves the highest AUC-ROC for Logistic Regression on Telco Churn, demonstrating competitive discrimination ability in a real-world imbalanced setting.

These findings establish QCBM as a viable, complementary tool for data augmentation—particularly suited to low-dimensional structured data—and motivate further investigation into more expressive circuit architectures, alternative encoding strategies, and hybrid quantum-classical generative models for broader applicability.

\section*{Acknowledgment}

The authors thank the Department of Computer Engineering, Faculty of Engineering, King Mongkut's University of Technology Thonburi for support and resources. We also thank Asst. Prof. Rajchawit Sarochawikasit for valuable co-advisory guidance.



\begin{thebibliography}{99}

\bibitem{goodfellow2016deep}
I.~J. Goodfellow, Y.~Bengio, and A.~Courville, \emph{Deep Learning}. MIT Press, 2016. [Online]. Available: \url{https://www.deeplearningbook.org}

\bibitem{chawla2002smote}
N.~V. Chawla, K.~W. Bowyer, L.~O. Hall, and W.~P. Kegelmeyer, ``SMOTE: Synthetic minority over-sampling technique,'' \emph{J. Artif. Intell. Res.}, vol.~16, pp.~321--357, 2002.

\bibitem{goodfellow2014generative}
I.~J. Goodfellow \textit{et al.}, ``Generative adversarial nets,'' in \emph{Adv. Neural Inf. Process. Syst.}, 2014, pp.~2672--2680.

\bibitem{biamonte2017quantum}
J.~Biamonte \textit{et al.}, ``Quantum machine learning,'' \emph{Nature}, vol.~549, no.~7671, pp.~195--202, 2017.

\bibitem{liu2018differentiable}
J.-G. Liu and L.~Wang, ``Differentiable learning of quantum circuit Born machine,'' \emph{Phys. Rev. Research}, vol.~2, no.~1, p.~013 284, 2020.

\bibitem{benedetti2019parameterized}
M.~Benedetti, E.~Lloyd, S.~Sack, and M.~Fiorentini, ``Parameterized quantum circuits as machine learning models,'' \emph{Quantum Sci. Technol.}, vol.~4, no.~4, p.~043 001, 2019.

\bibitem{gili2023quantum}
K.~Gili \textit{et al.}, ``Do quantum circuit Born machines generalize?'' \emph{arXiv:2207.13645}, 2022.

\bibitem{mohanty2025quantum}
N.~Mohanty, B.~K. Behera, C.~Ferrie, and P.~Dash, ``A quantum approach to synthetic minority oversampling technique (SMOTE),'' \emph{Quantum Mach. Intell.}, vol.~7, no.~1, p.~38, 2025.

\bibitem{han2005borderline}
H.~Han, W.-Y. Wang, and B.-H. Mao, ``Borderline-SMOTE: A new over-sampling method in imbalanced data sets learning,'' in \emph{Proc. ICIC}, 2005, pp.~878--887.

\bibitem{last2017oversampling}
M.~Last, A.~Litvak, and B.~B. Rokach, ``Data oversampling using KMeans cluster-based SMOTE,'' in \emph{Proc. ICPRAM}, 2017.

\bibitem{xu2019modeling}
L.~Xu, M.~Skoularidou, A.~Cuesta-Infante, and K.~Veeramachaneni, ``Modeling tabular data using conditional GAN,'' in \emph{Adv. Neural Inf. Process. Syst.}, 2019.

\bibitem{xu2018synthesizing}
L.~Xu and K.~Veeramachaneni, ``Synthesizing tabular data using generative adversarial networks,'' \emph{arXiv:1811.11264}, 2018.

\bibitem{jordon2019pate}
J.~Jordon, J.~Yoon, and M.~van der Schaar, ``PATE-GAN: Generating synthetic data with differential privacy guarantees,'' in \emph{Proc. ICLR}, 2019.

\bibitem{schuld2018supervised}
M.~Schuld and F.~Petruccione, \emph{Supervised Learning with Quantum Computers}. Springer, 2018.

\bibitem{mitarai2018quantum}
K.~Mitarai, M.~Negoro, M.~Kitagawa, and K.~Fujii, ``Quantum circuit learning,'' \emph{Phys. Rev. A}, vol.~98, no.~3, p.~032 309, 2018.

\bibitem{fisher1936use}
R.~A. Fisher, ``The use of multiple measurements in taxonomic problems,'' \emph{Ann. Eugenics}, vol.~7, no.~2, pp.~179--188, 1936.

\bibitem{ibm2018telco}
IBM Corp., ``Telco customer churn dataset,'' 2018. [Online]. Available: \url{https://www.ibm.com/communities/analytics/watson-analytics-blog/guide-to-sample-datasets/}

\bibitem{javadi2024qiskit}
A.~Javadi-Abhari \textit{et al.}, ``Quantum computing with Qiskit,'' \emph{arXiv:2405.08810}, 2024.

\bibitem{pedregosa2011sklearn}
F.~Pedregosa \textit{et al.}, ``Scikit-learn: Machine learning in Python,'' \emph{J. Mach. Learn. Res.}, vol.~12, pp.~2825--2830, 2011.

\bibitem{preskill2018quantum}
J.~Preskill, ``Quantum computing in the NISQ era and beyond,'' \emph{Quantum}, vol.~2, p.~79, 2018.

\bibitem{nielsen2010quantum}
M.~A. Nielsen and I.~L. Chuang, \emph{Quantum Computation and Quantum Information}. Cambridge Univ. Press, 2010.

\bibitem{esteban2017rgan}
C.~Esteban, S.~L. Hyland, and G.~R{\"a}tsch, ``Real-valued (medical) time series generation with recurrent conditional GANs,'' \emph{arXiv:1706.02633}, 2017.

\end{thebibliography}
\end{document}